\documentclass[runningheads]{llncs}
\usepackage{graphicx}
\usepackage{subfigure}

\usepackage{amsmath}

\begin{document}
\title{PF-Net: Personalized Filter for Speaker Recognition from Raw Waveform \thanks{This work was funded by the National Key Research and Development Program of China under Grant No. 2019YFB1405803.} }
%
%
%




\author{{Wencheng Li, Zhenhua Tan*, Zhenche Xia, Danke Wu, Jingyu Ning}
\institute{School of Software, Northeastern University, Shen Yang 110819, China\\
Email: liwencheng@stumail.neu.edu.cn, \{tanzh, ningjy\}@mail.neu.edu.cn, \{xiazhenche,wudk2019\}@stumail.neu.edu.cn \\
*Corresponding Author: tanzh@mail.neu.edu.cn}
}

%
\maketitle
\begin{abstract}
Speaker recognition using i-vector has been replaced by speaker recognition using deep learning. Speaker recognition based on Convolutional Neural Networks (CNNs) has been widely used in recent years, which learn low-level speech representations from raw waveforms. On this basis, a CNN architecture called SincNet proposes a kind of unique convolutional layer, which has achieved band-pass filters. Compared with standard CNNs, SincNet learns the low and high cut-off frequencies of each filter. This paper proposes an improved CNNs architecture called PF-Net, which encourages the first convolutional layer to implement more personalized filters than SincNet. PF-Net parameterizes the frequency domain shape and can realize band-pass filters by learning some deformation points in frequency domain. Compared with standard CNN, PF-Net can learn the characteristics of each filter. Compared with SincNet, PF-Net can learn more characteristic parameters, instead of only low and high cut-off frequencies. This provides a personalized filter bank for different tasks.  As a result, our experiments show that the PF-Net converges faster than standard CNN and performs better than SincNet. Our code is available at github.com/TAN-OpenLab/PF-NET.

\end{abstract}
\section{Introduction}
In our daily life, we are always receiving and conveying a variety of information from the outside world, and voice information is an important part of it. Audio signal processing plays an important role in the field of artificial intelligence and machine learning. Because each person's voice organs are different, their voices and tones are different. Apart from physical differences, each person has his own unique way of speaking, including using a specific accent, rhythm, intonation style, pronunciation mode, vocabulary selection, and so on. Together, these make up a distinct feature for everyone~\cite{1,2,3}.

In the early days, speaker recognition used models such as Dynamic Time Warping (DTW)~\cite{4} to calculate the similarity of speakers. Gaussian Mixture Model- Universal Background Models(GMM-UBM)~\cite{5,6,7} has improved on this basis. i-vector~\cite{8} reduces the computational cost by compressing the speaker vector and can make better use of the channel compensation algorithm. Most speaker recognition requires pre-processing steps. Incorrect pre-processing of recorded speech input can reduce classification performance~\cite{9}. Some common methods include noise removal, endpoint detection, pre-emphasis, framing, and normalization~\cite{10,11} can be used in pre-processing. After the rise of neural networks, researchers applied DNN~\cite{12} directly to speaker classification tasks. This method has the characteristics of large parameters and unclear timing information. The fully connected layer in d-vector was replaced by a feature extraction method similar to one-dimensional convolution in~\cite{13,14,15}, which reduces the computational cost. CNNs pays more attention to the speaker information between frames to achieve a better effect. Some networks used artificial features such as MFCC~\cite{9673167,9188632} and LPCC~\cite{23}. However, these features may cause the classifier to miss some speaker-specific information. In order to alleviate this shortcoming, some works used the spectrogram as the input of the network, the others directly used the raw waveform in their network. In order to process the raw waveform, a common choice for researchers is to apply CNNs. Because of the weight sharing feature, the convolution kernel can find a invariant and robust representation. In recent years, there have been many studies on speaker recognition based on the time-domain characteristics of audio using related neural networks~\cite{9053767,9251870}. Some studies have introduced ResNet~\cite{9420202,9383531} to deal with more and more complex network structures. However, they are not special in dealing with the first layer of neural network, and are usually the same as other layers in the network. SincNet~\cite{16} uses multiple sets of parameterized filters as the first layer of the convolutional network to obtain a personalized solution of the filter. However, SincNet uses the Sinc function as a filter, which only considers limited parameters.

RawNet~\cite{jung2019rawnet} mixes CNN with LSTM to get better performance. Its input layer is still the same as SincNet, which has the same limitation.
This paper alleviates the constraints of the Sinc function on the filter, and enables the filter to directly fit the shape in the frequency domain. This is formed by a set of parameterized linear filtering constraints. This method gives the filter more freedom and enables it to learn more peculiar features.

Our experiments were conducted under challenging but realistic conditions, which are characterized by few data (12-15 seconds per speaker) and short test sentences (lasting 2 to 6 seconds). The results show that the filter achieves better final performance.

The remainder of the paper is organized as follows. The PF-Net architecture is described in Sec. 2. Sec. 3 discusses the relation to prior work. The experimental setup and results are outlined in Sec. 4 and Sec. 5 respectively. Finally, Sec. 6 discusses our conclusions.



\section{The SincNet Architecture}
\label{sec:sinc}
For speech time series signals, the standard first-layer CNN structure is regarded as a time-domain convolution operation~\cite{17}, which is defined as follows:
\begin{equation}
y[n]=x[n]*h[n] = \sum\limits_{l=0}^{L-1} x[l]\cdot h[n-l]
\end{equation}
where $x[n]$ is the speech signal, $h[n]$ is the filter of length $L$, and $y[n]$ is the output of the filter. In the standard CNN structure, the $L$ elements of each filter are learned from the data.

In the SincNet structure, the convolution operation uses a predefined function $g$, where $g$ contains only a few learnable variables, defined as follows:
\begin{equation}
y[n]=x[n]*g[n,\theta]
\end{equation}
SincNet uses the low and high frequencies to filter the original audio in segments. The $g$ function is defined as a rectangular bandpass filter in the text, and its frequency domain characteristics are as follows:
\begin{equation}
G[f,f_{beg},f_{end}]= rect\Big(\frac{f}{2f_{end}}\Big) - rect\Big(\frac{f}{2f_{beg}}\Big)
\end{equation}
where $f_{beg}$ and $f_{end}$ are start cut-off frequency and end cut-off frequency respectively, both of which are learnable.

Unlike SincNet, the filter in PF-Net convolution operation reconstructs the function $g$ so that $g$ contains more learnable variables. This is achieved by inserting learnable deformation points in the waveform segments at different low and high frequencies, and divide a frequency domain into multiple one can learn small line segments. The frequency domain characteristics of each segment are expressed as follows:
\begin{equation}
G[f,f_{k},f_{k+1}]=\frac{h_{k+1}-h_{k}}{f_{k+1}-f_{k}}(f-f_{k})+h_{k}
\end{equation}
$f_{k}$ and $f_{k+1}$ represent the low and high frequencies of each segment, $h_{k}$ and $h_{k+1}$ are their corresponding amplitude intensity in frequency domain. $f_{k}$ , $f_{k+1}$, $h_{k}$ and $h_{k+1}$ are learnable, and the value of f is within $[f_{k},f_{k+1}]$.

When $f$ is not in the interval $[f_k,f_{k+1}]$:
\begin{equation}
G[f,f_{k},f_{k+1}]=0,
\end{equation}
Accumulate each line segment to get the whole frequency band that can be learned. The expression is as follows:
\begin{equation}
G[f,f_{beg},f_{end}] = \sum\limits_{k=beg}^{end-1} G[n,f_{k},f_{k+1}]
\end{equation}
After inverse Fourier transform~\cite{18}, the above formula about each line segment is transformed into the time domain expression form as:
\begin{equation}
\begin{split}
g[n,f_{k},f_{k+1}]=\frac{\Delta(cos(2\pi f_{k+1} n)-cos(2\pi f_{k} n))}{4\pi ^2 n^2}\\
-\frac{h_{k+1}sin(2\pi f_{k+1} n)-h_{k}sin(2\pi f_{k} n)}{2\pi n}
\end{split}
\end{equation}
where:
\begin{equation}
\Delta=\frac{h_{k+1}-h_k}{f_{k+1}-f_k}
\end{equation}

\begin{figure}
\centering
\includegraphics[width=0.50\textwidth]{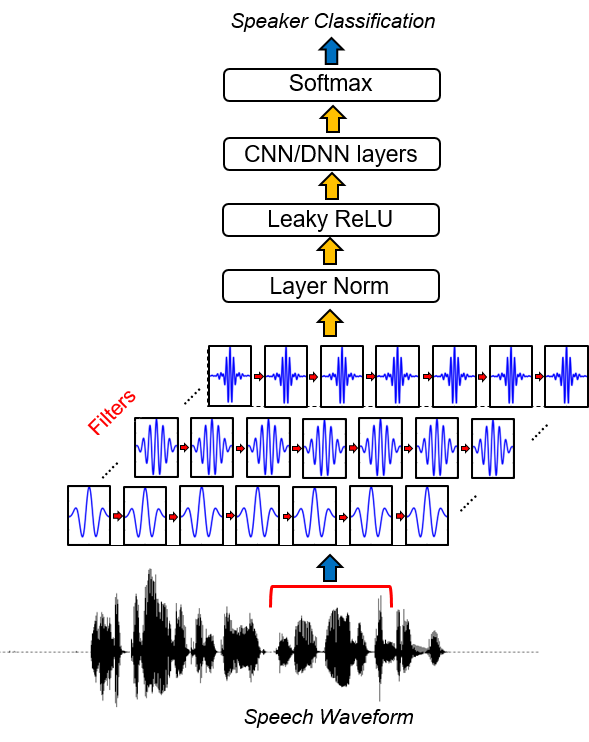}
\caption{Network Architecture of SincNet~\cite{16}, PF-Net differs from it only in the filter part} \label{fig:line_arch}
\end{figure}

\begin{figure}
\centering
\includegraphics[width=0.5\textwidth]{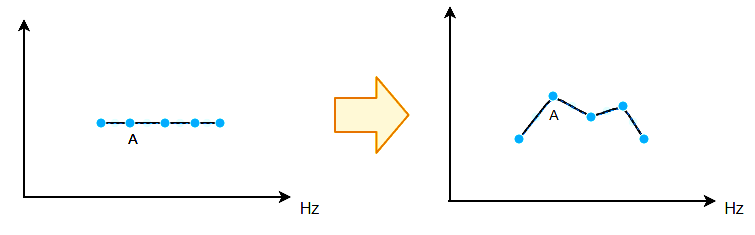}
\caption{Learning process of a filter in frequency domain.} \label{fig:filter}
\end{figure}

The time domain expression of the entire filtering frequency band is:
\begin{equation}
g[f,f_{beg},f_{end}] = \sum\limits_{k=beg}^{end-1} g[n,f_{k},f_{k+1}]
\end{equation}
Among them, the cutoff frequencies is initialized to a value in the range of $[0,f_s/2]$, and $f_s$ is the sampling rate of the signal. In fact, in order to better extract effective information from low frequencies, both the cutoff frequency and the deformation point frequency are initialized with Mel frequency.
To constrain the shape of the filter, $h_k$ is expressed as:
\begin{equation}
h_k=1+\Delta h
\end{equation}
By constraining the value of $\Delta h$, the value of h can be constrained indirectly. This paper uses $\Delta h$ instead of $h$ as the learnable parameters, so that $h_k$ always learns with 1 as the center. The value of $\Delta h$ is initialized randomly between $[-0.1,0.1]$.
In this way, the deep neural network can learn the shape of each filter. As shown in Fig \ref{fig:filter}, the initial shape of the filter in the frequency domain is the same as that of SincNet. After learning, the deformation points set in the frequency band begin to move in different directions by changing the abscissa and ordinate of the points (both of which are learnable parameters). Therefore, PF-Net can obtain richer information than the cut-off frequency.

Finally, in order to smooth the truncation characteristics of the $g$ function, multiply the $g$ function by a window function:
\begin{equation}
g_{w}[n,f_{beg},f_{end}]= g[n,f_{beg},f_{end}] \cdot w[n].
\end{equation}
Adopt Hamming window~\cite{19}:
\begin{equation}
w[n]= 0.54-0.46 \cdot cos\Big(\frac{2\pi n}{L}\Big).
\end{equation}
The neural network used in this paper is divided into two parts, the first part is the filter layer, the follow-up is the neural network part composed of CNN/DNN, and finally the posterior probability is output by Softmax. Preprocessing includes removing silence, framing, and maximum normalization operations. Among them, the first layer filter is the key design part, and the network structure is as Fig \ref{fig:line_arch}.

\subsection{Model properties}

\begin{itemize}
\item \textbf{Flexible parameters:}
PF-Net can control the number of parameters in the first convolutional layer. For example, if we consider a convolutional layer with F channels, each channel uses a filter of length L, and the number of deformation points is S, the standard CNN uses F$\cdot$L parameters, SincNet uses 2F parameters, and PF-Net uses L$\cdot$F$\cdot$S parameters, if F=80, L=200, S=5, CNN has 16000 parameters, and SincNet has 160. PF-Net has 800 parameters, somewhere in between. If you change the value of S, you can change the number of parameters without changing the number of channels. Moreover, if we filter twice the length L, twice the parameter count of the standard CNN, while the amount of parameters of PF-Net remains unchanged. This provides a very selective filter without actually adding parameters to the optimization problem. In addition, as the number of parameters increases, the PF-Net network has a more powerful fitting function, and at the same time, it is more difficult to train. The user can freely set the number of deformation points according to the data and actual needs.

\item \textbf{Interpretability:}
Same as SincNet, the features obtained by PF-Net in the first convolutional layer are more interpretable and readable than other methods. The reason for this property of PF-Net is that all deformation points have corresponding positions in frequency domain.
\end{itemize}

\section{Related Work} \label{sec:rel_work}
In the process of speaker recognition, feature extraction has attracted much attention. Researchers always assume that the input signal has a stable property in a short enough time interval. Therefore, segmenting the input signal into multiple short-term frames and extracting their feature sequences can better model the audio signal~\cite{20}. In the past, people mainly used manual features such as MFCC~\cite{9673167,9188632}, LPCC~\cite{23}, and PLP~\cite{24}. These methods will inevitably lose part of the speaker information. In the field of CNN, there have been many attempts to process audio using amplitude spectrograms~\cite{25,9465954,9726767,28,29}. Compared with manual features, the amplitude spectrogram retains more speaker features. However, this method requires careful adjustment of some key hyperparameters such as frame window length, overlap, and type. Therefore, it has become a trend to conduct deep learning directly through raw audio~\cite{31,32,33,34,35} has proved its feasibility.

SincNet~\cite{16} uses a set of parameterized Sinc filters to directly learn from the original audio. Liu proposed a learnable MFCC~\cite{9401593}, but only carried out experiments in speaker verification. In order to obtain the timing characteristics in the original audio, many neural networks ~\cite{9053767,9251870} used to process timing signals are used for speaker recognition. To deal with more complex neural networks, ResNet~\cite{9420202,9383531} was introduced into speaker recognition. These networks pay more attention to the deep structure, but pay less attention to the first layer structure of the network. RawNet~\cite{jung2019rawnet} mixes CNN with LSTM to get better performance, which has the same first layer as SincNet. However, for each filter, SincNet only has two parameters that can be learned.

In this paper, a neural network which called PF-Net is proposed to learn the shape of the filter by inserting deformation points into each filter. PF-Net can actively control the number of learnable parameters by controlling the number of deformation points. This kind of filter has a good effect on speaker recognition, especially in the scene with several seconds samples. Each speaker provides a short test statement.

\begin{figure}

\subfigure[CNN Filters]{\includegraphics[width=0.45\textwidth]{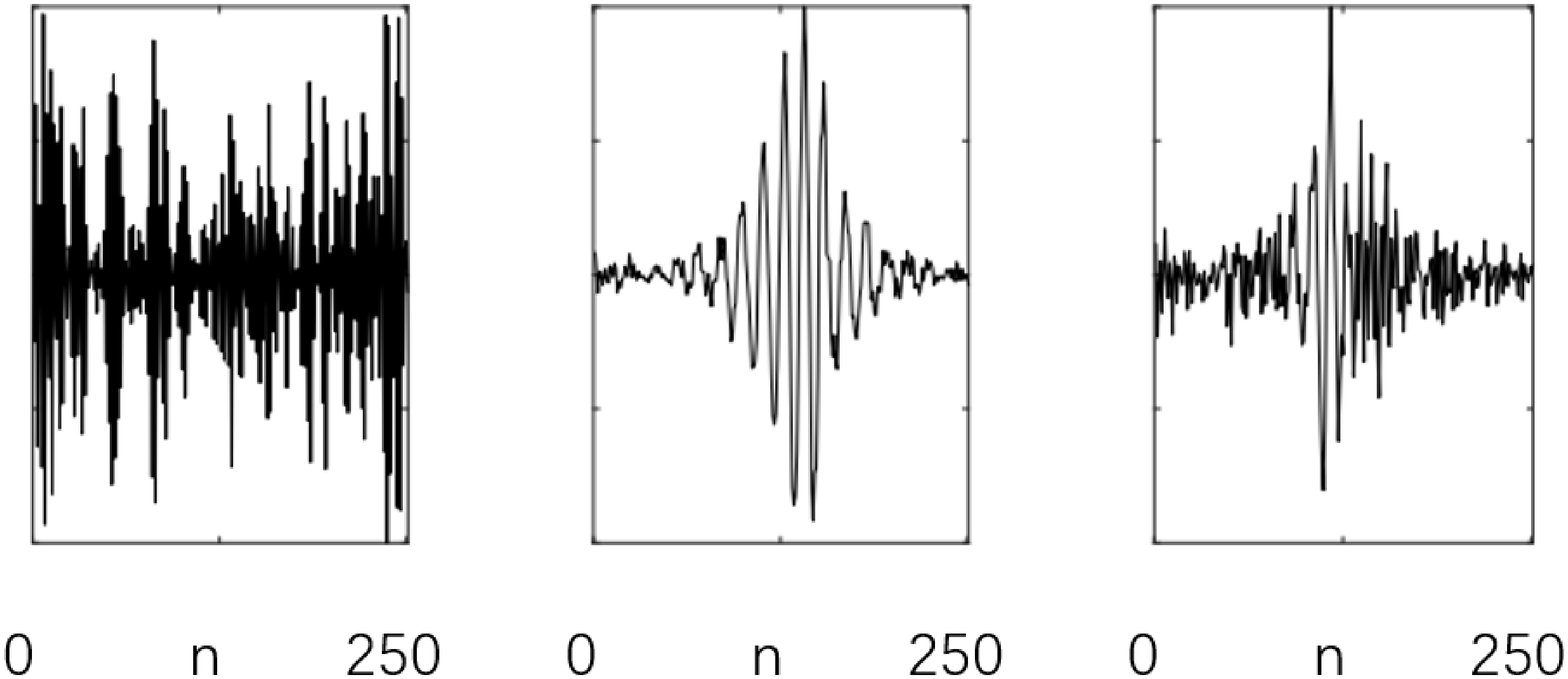}%
\label{fig:cnn_filt_time}}
\hfil
\subfigure[SincNet Filters]{\includegraphics[width=0.45\textwidth]{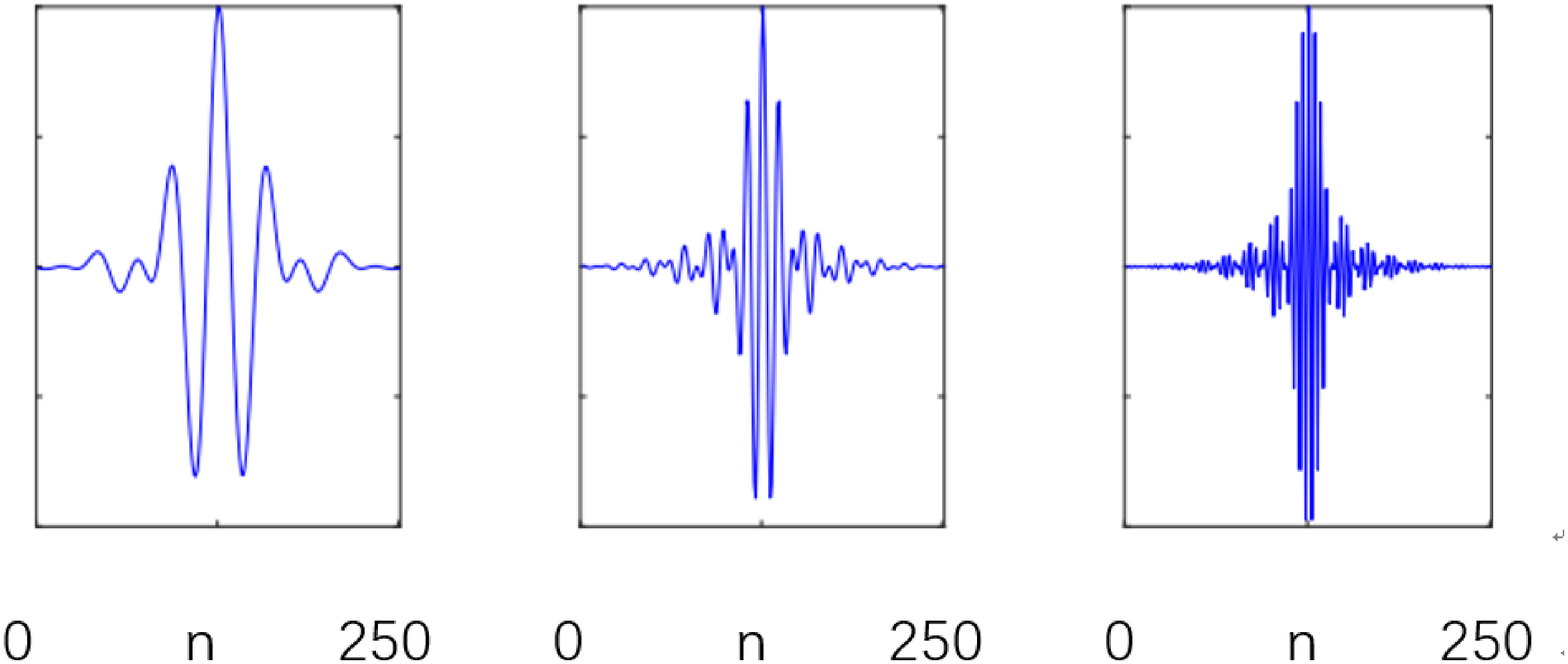}%
\label{fig:sinc_filt_time}}
\hfil
\centering
\subfigure[PF-Net Filters]{\includegraphics[width=0.45\textwidth]{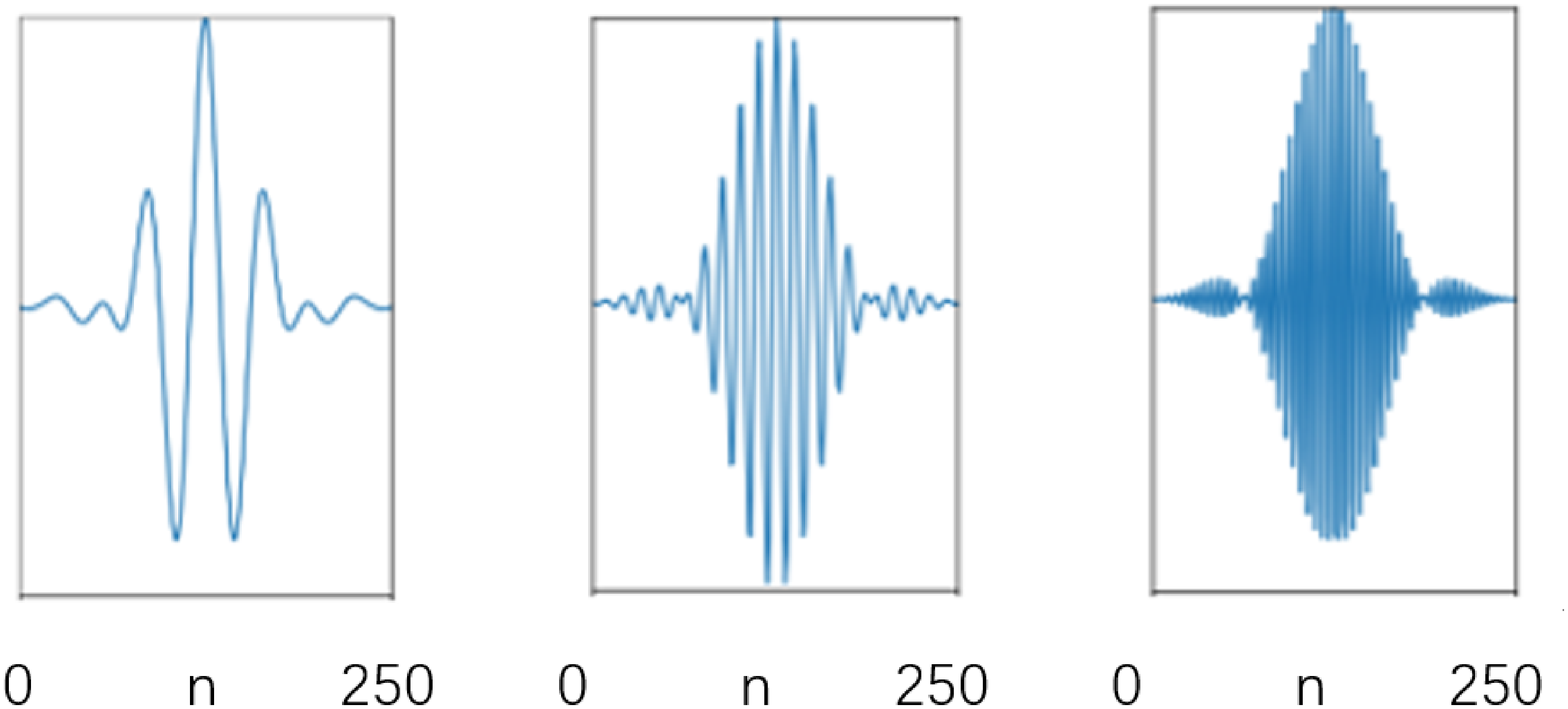}%
\label{fig:line_filt_time}}

\caption{Time domain representation of different channel filters earned by neural networks.}
\label{fig:filter_time}

\end{figure}

\begin{figure}

\subfigure[CNN Filters]{\includegraphics[width=0.45\textwidth]{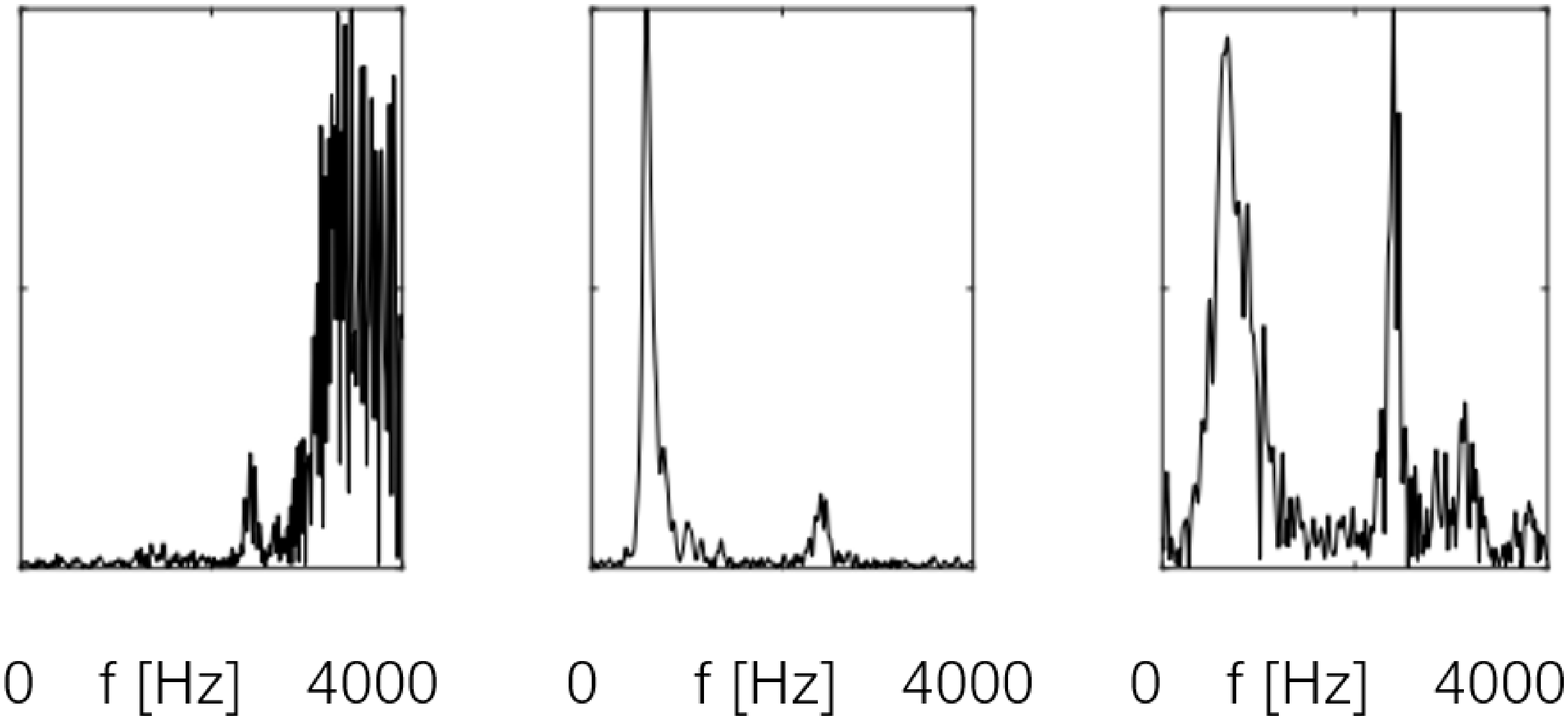}%
\label{fig:cnn_filt_fre}}
\hfil
\subfigure[SincNet Filters]{\includegraphics[width=0.43\textwidth]{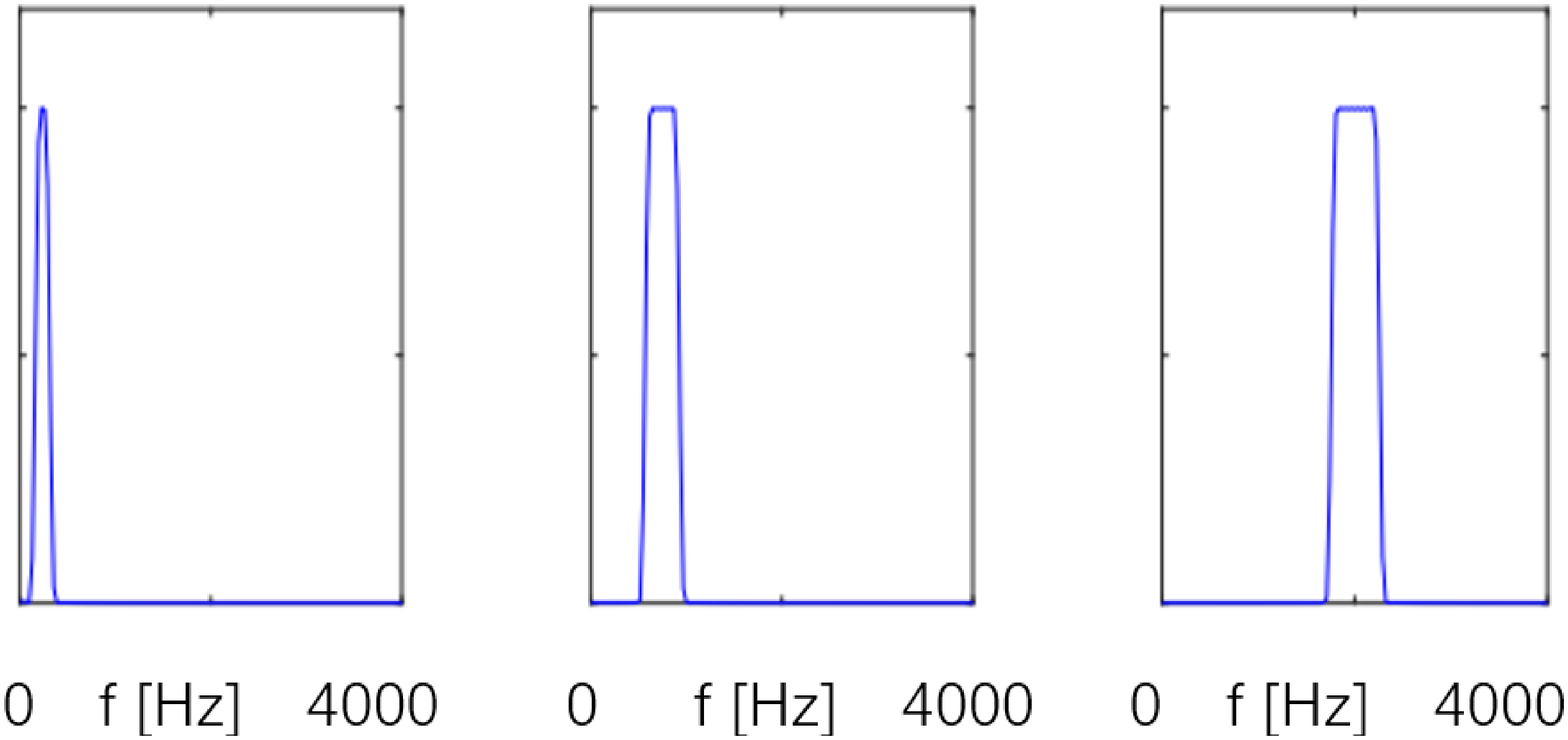}%
\label{fig:sinc_filt_fre}}
\hfil
\centering
\subfigure[PF-Net Filters]{\includegraphics[width=0.45\textwidth]{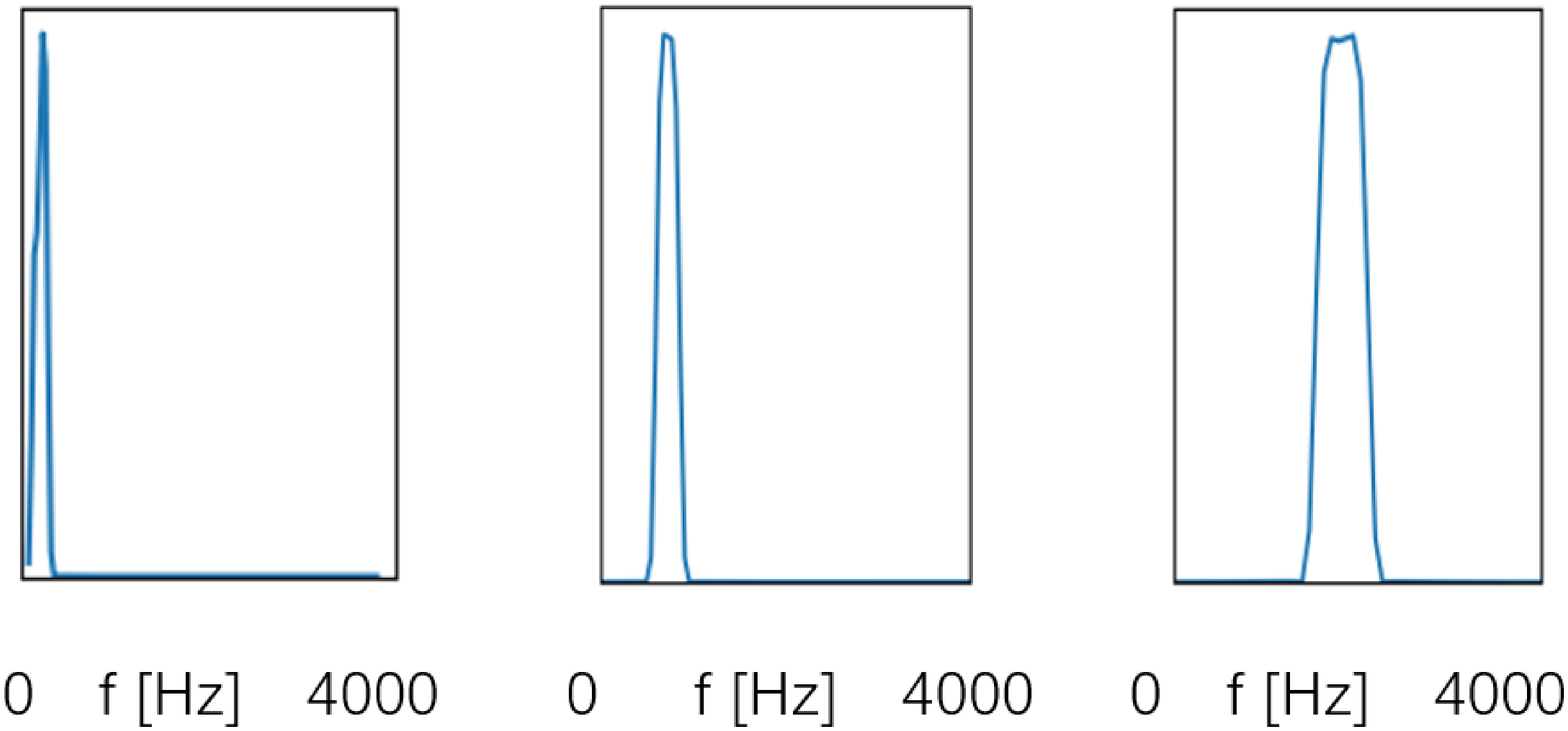}%
\label{fig:line_filt_fre}}
\caption{Time domain representation of different channel filters earned by neural networks.}
\label{fig:filter_fre}

\end{figure}

\section{Experimental Setup}
On some public datasets (TIMIT and Librispeech), PF-Net was compared with different baseline systems. In the following sections, we show the detailed settings of the experiment.
\label{sec:setup}
\subsection{Corpora}
This paper conducts experiments on the following two datasets.

TIMIT corpora~\cite{timit}, the voice sampling rate of the TIMIT dataset is 16kHz, which contains a total of 6300 sentences, and 70\% of the speakers are male. Most of the speakers are adult whites. Use the TIMIT dataset training set part for the speaker recognition task (same as SincNet), After deleting TIMIT's calibration sentences (same sentences), each speaker had 8 sentences, of which five sentences were used to train the model, and the remaining three were used for testing.

Librispeech corpora~\cite{librispeech}, which is a corpus of about 1000 hours of 16kHz reading English speech. It contains 2484 speakers, and the sampling rate is 16khz. The training and test speech in the dataset are used in the experiment. After removing the silent part, each sentence lasts for 2-6s.

\subsection{PF-Net Setup}
Speech signal is divided into frames (frame length 200$ms$, overlapping 10$ms$) and then input into PF-Net. We insert five deformation points in each filter. As Fig.2, The filter layer uses 80 filters of length 251, which are calculated by the formula in Section 2. Subsequently, the CNN part used two standard convolutional layers, the number of channels was 60, and the size of the convolution kernel was 5. Layer normalization~\cite{36} was used for input samples and for all convolutional layers (so is the filter layer). Next, we used 3 fully connected layers containing 2048 neurons, which normalized with batch normalization~\cite{37}. The activation function used leaky-ReLU~\cite{38}. The transformation point was initialized with mel-scale. We used softmax classifier to output the posterior probability of each frame. On this basis, the results of sentence-level classification were obtained by voting.
The RMSprop optimizer was applied, and the learning rate lr = 0.001, $\alpha$ = 0.95, $\varepsilon$ = $10^{-7}$. The batch size is 128. All hyper-parameters are optimized on TIMIT, as is Librispeech.The speaker verification system can directly take the softmax posterior score corresponding to the claimed identity~\cite{8462165}.

\subsection{Baseline Setups}
This paper compares PF-Net with several different systems.

First of all, CNN network with raw waveform as input was applied. Its network structure is the same as PF-Net in this paper, but the first layer convolution was replaced by PF-Net convolution. Secondly, we compared with manual features, this paper uses Kaldi toolkit~\cite{39} to calculate 40 FBANK features, which were calculated every 25 ms, with an overlap time of 10 ms, forming a context window of 200 ms. Layer normalization was also used in FBANK network and processed by CNN network. Finally, Sinc function was used to replace the first layer convolution of CNN. Except for the first layer, the network structure and hyperparameters of PF-Net were all the same as SincNet~\cite{16}. For speaker verification experiments, the enrollment and test phase is conducted on Librispeech.

\section{Results}
This section reports the experimental validation of the proposed SincNet. First, we perform a comparison between the filters learned by a SincNet and by a standard CNN. We then compare our architecture with other competitive systems on speaker recognition tasks.

\label{sec:exp}

\subsection{Filter Analysis}
Fig \ref{fig:filter_time} and Fig \ref{fig:filter_fre} are visual representations of the time-frequency domain of the filters learned by CNN, SincNet and PF-Net in the librispeech dataset, respectively. Through the analysis of them, we can find that PF-Net has the advantages of both original CNN and SincNet. Fig \ref{fig:filter_time} represents the time-domain waveforms of the three filters. Through observation, we can conclude that the filters learned by standard CNN are always noisy. In contrast, PF-Net and SincNet can learn more interpretable filters. In the time domain, In the time domain, SincNet found a narrower waveform while PF-Net found a wider one. In the frequency response shown in Fig \ref{fig:filter_time}, the standard CNN is more difficult to interpret, and the noise is almost distributed in all frequency bands. For SincNet structure, although the filter can learn the corresponding response of a specific frequency band, its model remains unchanged in amplitude intensity. Fig \ref{fig:line_filt_fre} shows that PF-Net is different from both. It not only realizes band-pass filtering, but also is not invariable in amplitude intensity.

\subsection{Speaker Identification}

\begin{table}

\caption{Classification Error Rates (CER\%) of PF-Net on two datasets}
\label{tab:spk_id_res}
\centering
\begin{tabular}{|l|l|l|}
\hline
\textbf{Model} & \textbf{TIMIT} &  \textbf{LibriSpeech}  \\
\hline
 CNN-FBANK           &   0.86      &     1.55     \\

 CNN-Raw              &   1.65      &    1.00     \\

 SINCNET                 &   0.85      &     0.96     \\

 PF-NET        &   \textbf{0.72}   &  \textbf{0.77}   \\
\hline
\end{tabular}
\end{table}

Table~\ref{tab:spk_id_res} shows the comparison of Classification Error Rates (CER\%). This table shows the comparison of PF-Net with other networks on the LibriSpeech dataset and TIMIT dataset. The network using CNN raw has a lower accuracy rate on the TIMIT dataset and a higher accuracy rate on the LibriSpeech dataset, which indicates the advantage of the original filter, that is, it is more suitable for datasets with large data volume, which is also a feature of the neural network. CNN network using fbank performs well on data sets with small data volume due to the use of manually extracted features. However, it is not as good as CNN on larger datasets. Sincnet and PF-net are both manually designed features and retain learnable parameters. Both of them combine the learnability of neural network and the experience of manual design. Therefore, in two datasets of different sizes, the CER can still be kept low. Among them, because PF-net has designed more parameters that can be learned, it can learn more complex filter banks, which makes PF-net outperform sincnet in datasets. This is because larger datasets need more complex filters for characterization. Due to the small amount of data in TIMIT, Sincnet is not very different from PF-net. However, with the increase of the amount of data, simple bandwidth learning cannot adapt to more complex data, which makes a greater gap between the two in LibriSpeech.

The LibriSpeech dataset was fast in the first 500 rounds. Although the speed slowed down thereafter, there was still a performance improvement, and the training ended in the 2900th round. The Frame Error Rate (FER\%) of LibriSpeech is 0.3, which is better than the result on the TIMIT dataset.

\subsection{Speaker Verification}
\begin{table}
\centering
\caption{Equal Error Rates (EER\%) of PF-Net on two Librispeech dataset.}\label{tab:spk_id_ver}
\begin{tabular}{|l|l|l|}
\hline
\textbf{Model} & \textbf{LibriSpeech}   \\
\hline
 CNN-FBANK           &   0.37           \\

 CNN-Raw              &   0.36           \\

 SINCNET                 &   0.32           \\

 PF-NET        &   \textbf{0.30}      \\
\hline
\end{tabular}
\end{table}

Table~\ref{tab:spk_id_ver} shows the comparison of Equal Error Rate (EER\%). This table shows the performance of the PF-Net on the LibriSpeech dataset. It can be seen that PF-Net performs better than both a CNN trained on standard FBANK coefficients and CNN trained or the raw waveform in speaker verification task. Although the performance of PF-Net is close to SincNet in this task, it still has certain advantages.Through the above experiments, it can be seen that PF-net improves the baseline in speaker identification more than in speaker verification, which may be because the speaker verification task is more prone to over fitting than speaker recognition.

\section{Conclusion}
In this paper, PF-Net is proposed, which is a neural architecture for directly processing waveform audio. Our model is a change of SincNet, which forms a loose constraint on the shape of the filter through effective parameterization and retains a certain order. Compared with SincNet, PF-Net has more freedom and can adjust the number of parameters freely. PF-Net has carried out extensive evaluation for challenging speaker recognition tasks, showing its performance advantages in speaker recognition. In addition to improving the performance, PF-Net and SincNet have the same advantages, that is, the convergence speed of standard CNN is improved, and the computational efficiency is improved by using the symmetry of the filter. In future work, we plan to evaluate PF-Net performance on datasets in other languages (such as Cn-Celeb) or in more challenging environments (such as VoxCeleb). Although only speaker recognition experiments are carried out in this paper, in theory, PF-Net, a time series processing method, should also work in other fields. Therefore, our future work will be extended to emotion recognition and singer recognition.

\bibliographystyle{splncs00}
\bibliography{custom}

\end{document}